\documentclass[useAMS,usenatbib]{mn2e}
\usepackage{graphicx}
\usepackage{natbib}
\usepackage{amssymb}
\bibpunct{(}{)}{;}{a}{}{,}
\newcommand{\aap}{A\&A}

\newcommand{\apj}{ApJ}
\newcommand{\apjl}{ApJ}
\newcommand{\apjs}{ApJS}
\newcommand{\mnras}{MNRAS}
\newcommand{\aj}{AJ}
\newcommand{\araa}{ARA\&A}
\newcommand{\gca}{Geocosmochimica Acta\ }

\newcommand{\pasj}{PASJ\ }

\def\zw{I\,Zw\,1}

\def\xmm{{\it XMM-Newton}}
\def\epi{EPIC-pn}
\def\rg{RGS}

\def\Halpha{\ifmmode {\rm H}\alpha \else H$\alpha$\fi}
\def\Hbeta{\ifmmode {\rm H}\beta \else H$\beta$\fi}
\def\Hgamma{\ifmmode {\rm H}\gamma \else H$\gamma$\fi}
\def\Hdelta{\ifmmode {\rm H}\delta \else H$\delta$\fi}
\def\Lya{\ifmmode {\rm Ly}\alpha \else Ly$\alpha$\fi}
\def\Lyb{\ifmmode {\rm Ly}\beta \else Ly$\beta$\fi}
\def\Lyg{\ifmmode {\rm Ly}\beta \else Ly$\gamma$\fi}

\def\feiii{Fe\,{\sc iii}}

\def\hi{H\,{\sc i}}
\def\hii{H\,{\sc ii}}

\def\ciii{\ifmmode {\rm C}\,{\sc iii} \else C\,{\sc iii}\fi}
\def\civ{\ifmmode {\rm C}\,{\sc iv} \else C\,{\sc iv}\fi}
\def\cv{\ifmmode {\rm C}\,{\sc v} \else C\,{\sc v}\fi}
\def\cvi{\ifmmode {\rm C}\,{\sc vi} \else C\,{\sc vi}\fi}

\def\nv{N\,{\sc v}}

\def\oi{O\,{\sc i}}

\def\o5007{[O\,{\sc iii}]\,$\lambda5007$}

\def\ovi{O\,{\sc vi}}
\def\ovii{O\,{\sc vii}}
\def\oviii{O\,{\sc viii}}

\def\feiii{Fe\,{\sc iii}}
\def\fev{Fe\,{\sc v}}

\def\fexiv{Fe\,{\sc xiv}}
\def\fexvi{Fe\,{\sc xvi}}

\def\o{\o}
\def\kms{km\,s$^{-1}$}

\title[The ionised X-ray absorber of \zw\ observed by \xmm]{A longer \xmm\ look at I\,Zwicky\,1:\\ 
physical conditions and variability of the ionised
absorbers}

\author[E.~Costantini et al.]{E.~Costantini $^{1,2}$\thanks{E-mail: e.costantini@sron.nl}, L.C.~Gallo$^{3,4}$, 
W.N.~Brandt$^{5}$, A.C.~Fabian$^{6}$ and Th.~Boller$^{4}$\\
$^{1}$ SRON National Institute for Space Research, Sorbonnelaan 2, 3584 CA Utrecht, The Netherlands\\
$^{2}$ Astronomical Institute, Utrecht University, P.O. Box 80000, 3508 TA, Utrecht, The Netherlands\\
$^{3}$ SUPA, School of Physics and Astronomy, University of St.~Andrews, North Haugh, St.~Andrews, Fife KY16 9SS\\
$^{4}$ Max-Planck Institute f\"uer extraterrestrische Physik, Giessenbachstr 1, 45748, Garching bei Muenchen, Germany \\
$^{5}$ Department of Astronomy and Astrophysics, Pennsylvania State University, University Park, PA 16802\\
$^{6}$ Institute of Astronomy, University of Cambridge, Madingley Road, Cambridge CB3 0HA}

\begin{document}
\date{Received  / Accepted  }
\pagerange{\pageref{firstpage}--\pageref{lastpage}} \pubyear{2007}

\maketitle

\begin{abstract}
We present a spectral analysis of the narrow-line Seyfert~1 galaxy I~Zwicky\,1, focusing on the characteristics of the ionised absorbers as observed with \xmm\ in 2005. The soft X-ray spectrum shows
absorption by two components of ionised gas with a similar column density 
($N_{\rm H}\sim10^{21}$\,cm$^{-2}$) and ionisation parameters 
$\log\xi\sim0$ and $2.5$. Comparing this observation with a 2002 \xmm\ data set, we see a clear
anti-correlation between the X-ray ionisation parameter $\xi_{\rm X}$ and the $0.1-10$\,keV luminosity. Viable
explanations for this effect include transient clouds or filaments crossing the line of sight in a complex geometry or a 
gas observed in non-equilibrium. The outflow velocity of
the X-ray low-ionisation absorber is consistent with the outflow of the UV absorber detected in a past
Hubble Space Telescope observation. In
addition, the ionic column densities of \civ\ and \nv\ derived from the X-ray model are consistent with the UV values.
This suggests that the low-ionisation outflowing gas may survive for many years, despite large changes in flux, 
and that there is a tight connection between the X-ray and UV absorbers that can only be confirmed with a simultaneous UV and X-ray
observation.      
\end{abstract}


\section{Introduction}
About 50--70\% of active galactic nuclei (AGN) display absorption by photoionised gas in their 
X-ray spectra \citep[e.g.][]{reynolds97,george98}. With the advent of high energy resolution instruments, 
the study of the brightest sources has allowed us to detail the 
characteristics of such gas \citep[e.g.][]{kaastra02,kaspi02,kraemer05,turner05}. 
Multiple components are often detected, differing in column density, ionisation and outflow velocity. At least some of the 
lower ionisation X-ray components most probably share the same location as the UV components 
\citep[e.g.][]{turner05,elisa06} as first
suggested by \citet{smita95}. Variability studies point out the complex response of the 
ionised absorbers (also called
warm absorbers) to central-source flux variations. A linear response of the 
ionisation parameter\footnote{The ionisation parameter is defined as $\xi=L/nr^2$, where $L$ is the 1--1000 Ryd ionising 
luminosity, $n$ and $r$ are the
density and the distance to the ionised gas.} of the gas component to the continuum change provides a constraint on the 
distance to the absorber 
\citep[e.g.][]{netzer02,turner05,otani96}. A linear response gives a relatively intuitive picture of the geometry of the 
absorber as a thin slab of material in thermal equilibrium (either in clouds or filaments) exposed to the ionising flux. 
In other variability studies \citep[e.g. NGC~4151,][]{kraemer05} the gas does not obey this simple picture, 
suggesting instead
that gas outflows often change direction along the line of sight in a unpredictable way.\\    
I~Zwicky\,1 (\zw) is a well-studied narrow-line Seyfert~1 galaxy \citep[NLS1; see][and references therein]{vest_wilkes} 
located at
redshift $z$=0.06112$\pm$0.000015 \citep{condon}.
Since the time of ROSAT observations, \zw\ has been known to host intrinsic absorption in the X-ray band \citep{boller96,lawrence97}.
From moderate-resolution ASCA data, \citet{leighly99} could distinguish between a neutral component ($N_{\rm
H}\sim2.6\times10^{20}$\,cm$^{-2}$) and an ionised absorber, typically highlighted, in lower resolution
spectra, by 
the \ovii\ and \oviii\ absorption edges. A similar 
spectral complexity was reported by \citet{luigi04}, using \xmm\ data. The presence of fast ($v_{\rm
out}\sim-1870$\,\kms) outflowing
ionised gas was observed in the UV spectrum of \zw\ \citep[using HST-FOS;][]{laor97,crenshaw99}, 
through the analysis of the \nv\ and \civ\ absorption troughs. FUSE observed the \ovi\ absorption troughs, 
which displayed multiple components \citep[][]{kriss02}.\\      
This paper is one in a series devoted to a long \xmm\ observation of \zw. 
In the present study we focus
on the soft X-ray absorbers, discussing the physical picture returned by the long-term variability 
of this gas and its connection with the UV absorbers. Other works deal with the continuum emission and the variable
iron K$\alpha$ line (Gallo et al. 2007a, hereinafter Paper~I), and with the spectral variability analysis (Gallo et al. 2007b).

The paper is organised as follows. In Sect.~2 the main features of the data reduction and 
the \epi\ analysis 
(described in detail in Paper~I) 
are summarised. Sect.~3 is devoted to the analysis of the \rg\ data and the modelling of the 
spectral features in \zw. In Sect.~4 we discuss our results and in
Sect.~5 we present the conclusions. 
The cosmological parameters used are $H_0$=70 \kms Mpc$^{-1}$, $\Omega_m$=0.3, and $\Omega_{\Lambda}$=0.7.
The quoted errors are at 90\% confidence for one interesting parameter, unless
otherwise stated. The assumed Galactic
column density is $N_{\rm H}=5.07\times10^{20}$\,cm$^{-2}$ \citep{elvis89}.


\section{The data analysis}\label{sect:data}
\zw\ was observed on July 18-19 2005 (rev. 1027) for 85.5\,ks. 
Data from the three EPIC cameras MOS \citep{2001A&A...365L..27T} and pn \citep{2001A&A...365L..18S}, the
high resolution spectrometer RGS \citep{2001A&A...365L...7D} and the optical monitor \citep[OM, ][]{mason} 
were
successfully gathered. After filtering high-background intervals, the net exposure time was $\sim58$\,ks for
\epi\ and $\sim84$\,ks for RGS. The event files for each instrument were created with the newest SAS pipeline (ver.~7.0).
This version of SAS benefits from improved calibrations of \rg\ time-dependent effective area, absolute \rg\
flux and bad-pixel location\footnote{For further details refer to \\
http://xmm.vilspa.esa.es/external/xmm\_user\_support/usersgroup\\/20060518/index
.shtml.}. The discrepancies between \epi\ and \rg\ are of the
order of 10\%, similar as for EPIC-MOS when compared to \epi.\\ 
The spectral analysis of the soft X-ray spectrum was carried out using the fitting
package 
SPEX\footnote{http://www.sron.nl/divisions/hea/spex/version2.0/\\release/index.html} (ver.~2.0). 
We rebinned the \rg\ data to reach
a signal-to-noise ratio (S/N) of $\sim$9 over the whole energy band. During the observation the source shows two different average flux states, which differ approximately 
by 44\% and 36\% in the soft (0.3--2\,keV) and hard (2--10\,keV) energy bands, respectively. 
The flux changes after $\sim$27\,ks since the start of the observation and happens on a dynamical time scale of $\sim$5\,ks. 
The flux variations in the soft and hard X-ray bands show similar trends (see Fig.~3 of Paper~I). We will study the effect of the 
flux change on the RGS spectrum in Sect.~\ref{par:data}.\\ 
The data from the EPIC-pn camera were used to constrain the broad-band spectrum. The modelling of the continuum
is described in full detail in Paper~I. 
For the purpose of studying any narrow absorption/emission features in the \rg\ band, the detailed broad-band continuum shape is not
crucial. 
For simplicity, here we describe the continuum with a broken power-law, which fits the data well 
(Tab.\ref{t:rgs_05}). However in Sect.~\ref{par:conti}, we
test our results for different continua.


\section{The soft energy spectrum}\label{par:data}
\begin{figure}
\begin{center}

\resizebox{\hsize}{!}{\includegraphics[angle=90]{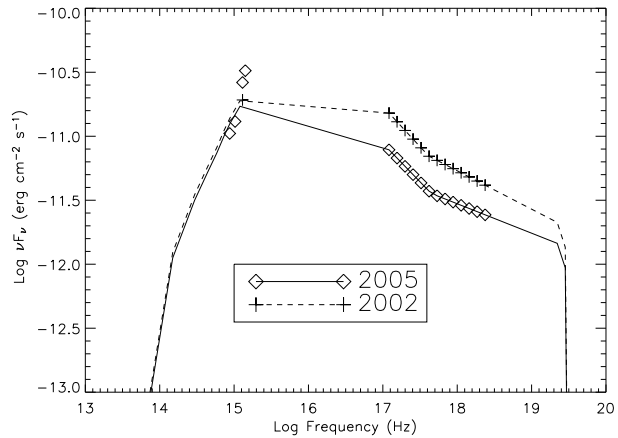}}
\end{center}
\caption{\label{f:sed} The adopted SEDs for the 05 observation (solid line) and 02 observation (dashed line). 
The X-ray shape of the SED is based on
the \epi\ data, while the point at 2500\AA\ is derived by interpolation of
the OM data points. The 05 OM points (open diamonds around 10$^{15}$\,Hz) were
interpolated to the same wavelength measurement of 02, in order to compare consistently the 05 SED with the 02 one, for
which data from only one OM filter was available.} 
\end{figure}

\begin{figure*}
\begin{center}

\resizebox{\hsize}{!}{\includegraphics[angle=90]{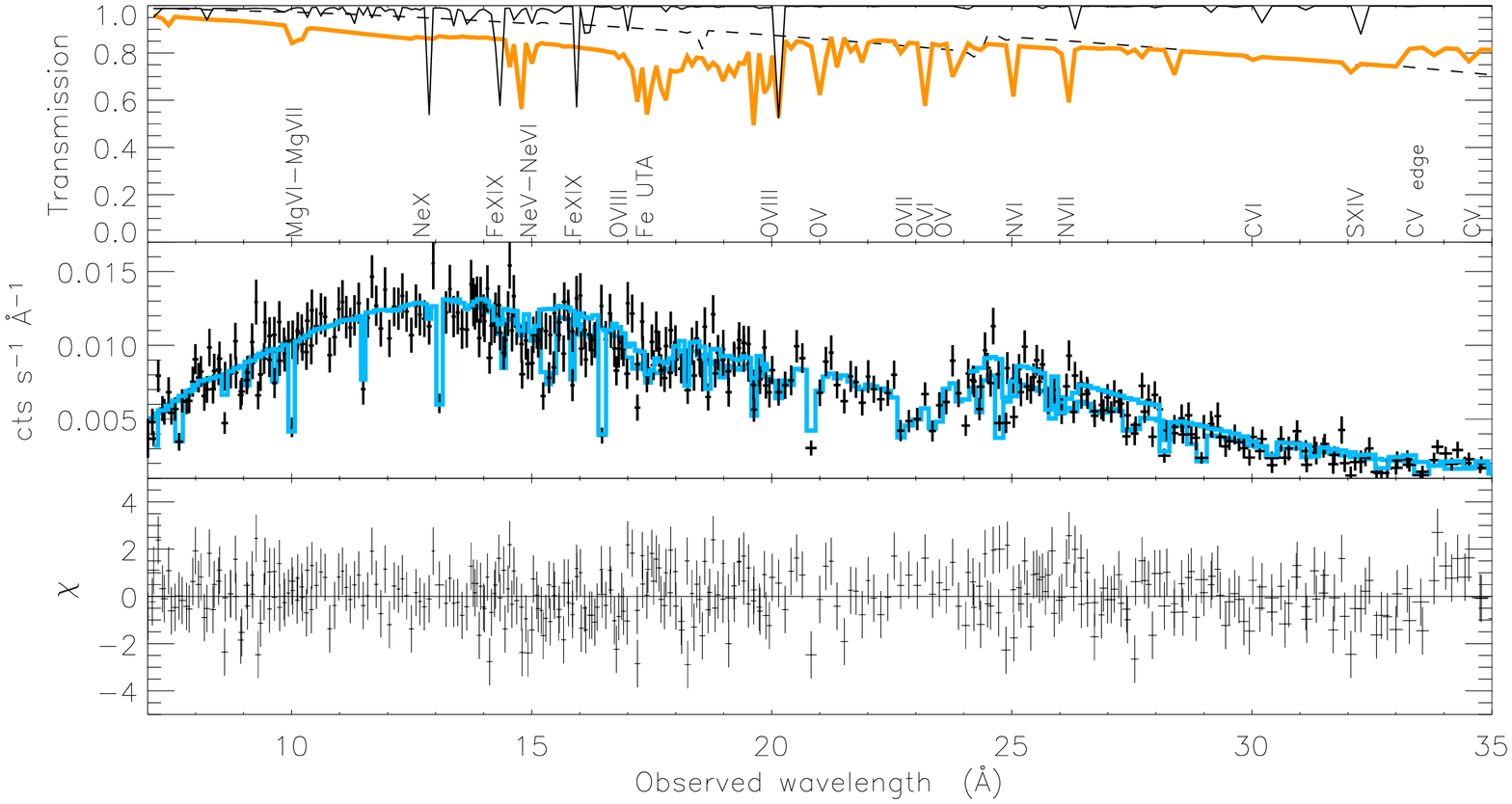}}
\end{center}
\caption{\label{f:rgs_tra05} Upper panel: 
transmission models (convolved with the data resolution) 
of the neutral absorber (dashed line) and the ionised components for the 05 spectrum. 
Component 05a is displayed with the 
light line and 05b with the dark line. The most relevant transitions are also labelled. 
Middle panel: \rg\ data together
with the best fit. Lower panel: residuals from the best-fit model.} 
\end{figure*}

The \rg\ spectrum shows the signatures of absorption, especially around the \ovii\ and \oviii\ absorption edges. 
In order to fit a photoionised absorber, we used the SPEX model XABS, which interpolates on a fine grid the $N_{\rm
H}$ and
$\log\xi$ values. The ionisation balance is constructed 
from the spectral energy distribution (SED) specific for the source (solid curve in Fig.~\ref{f:sed}), using Cloudy \citep{fer98}. 
Here and in the following analysis, the SED definition takes
advantage of the simultaneous \epi\ and OM data. The OM data were collected in fast-mode (see Paper~1) 
in the {\it U} (3440\AA), UVW1 (2910\AA), UVW2 (2120\AA) and UVM2 (2310\AA) filters. 
We considered the average flux for each filter.  
By
interpolation, we obtained the flux at 2500\,\AA\ that we used in constructing the SED. 
Interpolating the OM points was mandatory in order to 
compare consistently the ionisation balance obtained with this SED with the one obtained from a 
previous \xmm\ data set for which only one OM filter was available  (Sect.~\ref{par:xmm02}). An adopted SED will 
always have a degree
of uncertainty, when dealing with limited multiwavelength data. When comparing SEDs of different flux states of the same object it is important to consistently compare the resulting ionisation balances. The SED of 2002 was likely 
similar to 2005 in the UV regime, but in 2002 only one OM filter was available. 
This would make any ``reconstruction" of the 2002 SED highly arbitrary. Therefore the only way
to consistently compare the SEDs at the two epochs is to interpolate the 2005 OM points. Beyond the optical band, we used a standard AGN continuum, characterised by relatively low emission at longer wavelengths, as defined in 
Cloudy. The host galaxy of \zw\ is indeed very luminous both in the far-infrared and in the infrared \citep{halpern87}. 
In the long-wavelength band it is important to avoid dust contamination of the nuclear emission by the host galaxy. In general, an overestimation of the infrared radiation may lead to an incorrect ionisation balance determination \citep[see e.g.][]{ferl02}.\\ 
Tunable parameters of the XABS model, in addition to $N_{\rm H}$ and $\log\xi$, 
are the root mean squared width of the absorption
lines, the outflow velocity, the line-of-sight covering factor of the gas 
and the elemental abundances. We fixed the covering 
factor to unity and we assumed 
solar abundances \citep{anders} throughout the analysis.\\  
In Tab.~\ref{t:rgs_05} we list the best-fit parameters for the absorbers in \zw\ as measured by the RGS and, 
for consistency, with the lower resolution \epi\ data. 
The model parameters are always consistent between \rg\ and \epi. However, the higher energy resolution makes \rg\ more
reliable for the modelling of narrow absorption features, and we will consider the physical 
parameters constrained with the \rg\ data for
the discussion.\\  
A warm absorber (hereinafter labelled 05a), with a column
density $N_{\rm H}\sim1.3\times10^{21}$\,cm$^{-2}$ and an ionisation parameter $\log\xi\sim0$ 
is highlighted in the RGS data especially by the blend of the \ovii\ K absorption edge and the 
iron unresolved transition array (UTA). The gas is found to be outflowing at 
$\sim$1700\,\kms. The widths of the lines are loosely constrained ($\sigma=36\pm20$\,\kms), but still consistent with the 
more accurate FUSE measurement of 30$\pm$5\,\kms\ (G.~Kriss, private communication). 
Additional ionised absorption is required to fit the region below the 
\oviii\ edge ($\lambda=14.22$\AA). 
The improvement of the fit after the inclusion of this second absorber is $\Delta\chi^2/\Delta\nu$=27/2, where $\nu$ is the number
of degrees of freedom. The second gas component (hereinafter 05b) has a higher ionisation parameter ($\log\xi\sim2.6$,
Tab.~\ref{t:rgs_05}) but a column
density similar to the 05a component. We fixed the widths of the lines to the value for the lower ionisation
component. The absorption lines of the 
higher ionisation component lie in a spectral region of lower resolution, and a possible outflow velocity is hard to
quantify. The upper limit on the outflow velocity of the 05b component is
$v_{\rm out}<800$\,\kms. 
Finally, a neutral 
component, identified by an \oi\ edge at the redshift of the source and with $N_{\rm H}\sim1.5\times10^{20}$\,cm$^{-2}$, 
is required by the data. The addition of this component 
improves the fit by $\Delta\chi^2/\Delta\nu$=19/1 (where the column density is the additional parameter). 
In Fig.~\ref{f:rgs_tra05}, we plot the transmission of the neutral and warm
gas components (upper panel), the observed \rg\ spectrum with the best-fit model (middle panel) and the residuals from the
best fit (lower panel).\\ 
The above analysis refers to the total \rg\ spectrum. 
However (Sect.~\ref{sect:data}), 
the soft and hard energy bands undergo variations during the \xmm\ pointing. 
Using the \rg\ data we further verified that only continuum changes 
are responsible for the 
overall variations in the source flux and not opacity variations of the ionised gas. 
We divided the \rg\ data into two time segments,
corresponding to the higher ($\sim$30\,ks long) and lower state ($\sim$40\,ks). 
The higher state best fit does 
not vary significantly with respect to the average spectral fit, except for the normalisation and slope of the power law.    
We then used the higher flux state best-fit model (see above) as a template for the lower flux state and found that the
neutral and warm absorbers do not vary within the uncertainties. 

\begin{table}
\caption{\label{t:rgs_05}2005 best-fit model parameters for RGS and PN. 
Fluxes are in units of 10$^{-12}$ 
erg\,s$^{-1}$\,cm$^{-2}$, unabsorbed 
luminosity in units of 10$^{43}$ erg\,s$^{-1}$. Component 1: broken power law, 
2: neutral absorber, 3,4: warm absorbers. All emission components are modified by Galactic photoelectric absorption.}
\begin{center}
\begin{tabular}{llll}
\hline
comp. & param. &  R1+R2 & PN\\

1& $\Gamma_1$ & 2.5$\pm0.1$ & 2.58$\pm0.05$\\
 & $\Gamma_2$ & $\dots$ & 2.21$\pm$0.05\\
 & $E_{\rm break}$\,(keV) & $\dots$ & 2.0$\pm$0.1\\

2 & $N_{\rm H}$ (10$^{20}$ cm$^{-2}$) & 1.7$\pm$1& 1.5$\pm$0.4\\

3 & $N_{\rm H}$ (10$^{20}$ cm$^{-2}$) & 13$\pm3$ & 9$\pm1$\\ 
& $\log \xi$ & 0.05$\pm0.16$ & 0.16$\pm0.14$\\
& $v_{\rm out}$ (\kms) & --1700$\pm400$ & --1700 fix.\\

4 & $N_{\rm H}$ (10$^{20}$ cm$^{-2}$) & 13$\pm5$ & 8$\pm3$\\ 
& $\log \xi$ & 2.6$\pm0.3$ & 2.0$\pm0.4$\\
\hline
& $\chi^2/\nu$ & 411/358 &530/513\\

&  F$_{2-10\,{\rm keV}}$& $\dots$ &  4.86$\pm0.06$ \\
&  F$_{0.5-2\,{\rm keV}}$ &  5.0$\pm$0.3  & 5.13$\pm0.07$ \\
&  L$_{2-10\,{\rm keV}}$ &  $\dots$ & 4.50$\pm0.06$\\
&  L$_{0.5-2\,{\rm keV}}$&  7.0$\pm0.4$ & 6.10$\pm0.08$\\
\hline
\end{tabular}
\end{center}
\end{table}

\subsection{Other continuum models}\label{par:conti}

In the present paper the continuum is represented by a simple broken power law.
A discussion of the continuum models and
their physical implications is given in Paper~I. Here we test different continuum
models only 
in order to assess the robustness
of the warm-absorber parameters. 
We follow the same approach as in Sect.~\ref{par:data}, the broad-band continuum is established 
using \epi\ and then applied to the \rg, in order to tune the warm absorbers' parameters.\\
First we used a power law plus modified black-body emission and neutral absorption 
(both from the host galaxy and from the
Milky Way). In the black-body model the effects of Compton
scattering are considered \citep{kaastrabarr}. 
A black-body temperature of $\sim$0.13\,keV 
is required to fit the low energies. Without any warm absorber the resulting reduced $\chi^2$ ($\chi^2/\nu$) is $\sim$1.32. 
The addition of one warm absorber (two additional free parameters) improves the fit by $\Delta\chi^2$=93. 
A second component improves the fit by another $\Delta\chi^2/\Delta\nu=23/2$. 
The final model has a higher black body temperature, $kT\sim$0.2\,keV (see Paper~I for a discussion). 
The total black-body
luminosity is however very low ($\sim 2.7\times10^{42}$\,erg\,s$^{-1}$) compared to the dominant 
power law component ($\sim1.3\times10^{43}$\,erg\,s$^{-1}$) in the 0.3--10\,keV band. 
The warm absorbers parameters do not differ within the errors from the ones listed in Tab.~\ref{t:rgs_05}.\\  
Next, we tested a relativistically blurred reflection model \citep{ross05}. 
We fixed the parameters to the model values determined
in Paper~I, i.e. the disc inclination to 51$^{\circ}$, its emissivity
$\alpha=2.5$ 
and inner radius to 5
gravitational radii. We assumed solar abundances for the disc material. 
A fit without any ionised absorption results in a $\chi^2/\nu$ of $1.56$. This value decreases to $1.21$, 
adding one warm absorber. A second warm absorber component again improves the fit, as in the previous cases ($\Delta\chi^2/\Delta\nu=20/2$). 
Note that for this exercise we do not attempt any
fit of the Fe K$\alpha$ line. The amount of reflected light with respect to
the total luminosity budget is modest (few percent), resulting in a weak soft excess. 
We conclude that for the purpose of our study a simple broken-power law fit is acceptable.
\begin{figure}
\resizebox{\hsize}{!}{\includegraphics[angle=90]{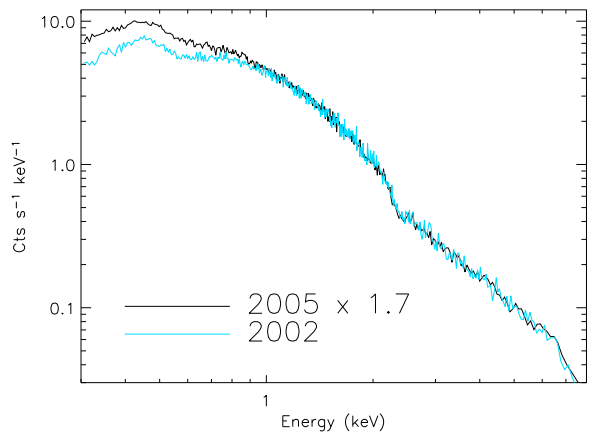}}
\caption{\epi\ spectra taken in 02 (light line) and 05 (dark line). The spectrum of 05 is multiplied by 1.7 to best match
the hard continuum during 2002 and emphasises the low energy complex variability.}
\label{f:ratio0205}
\end{figure}
\subsection{The recent history of the absorbed X-ray spectrum}\label{par:xmm02}
\begin{figure*}
\begin{center}
\resizebox{\hsize}{!}{\includegraphics[angle=90]{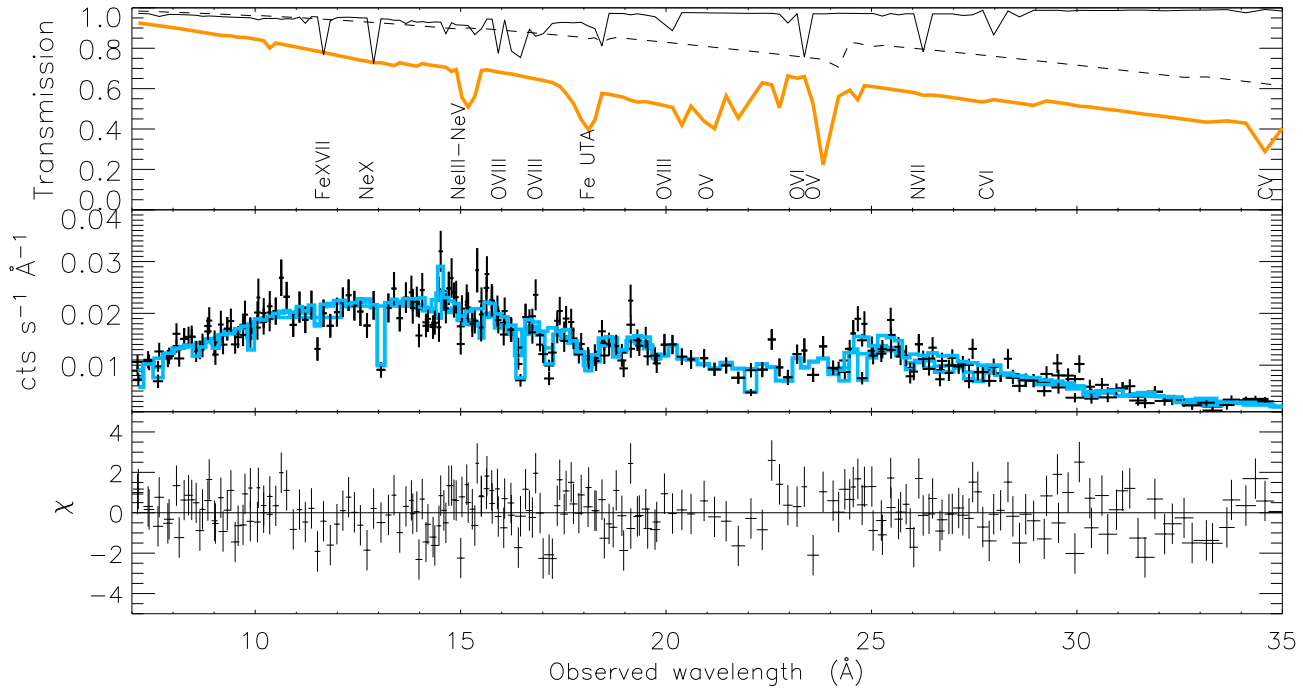}}
\end{center}
\caption{\label{f:rgs_tra02} Upper panel: 
transmission models (convolved with the data resolution)
of the neutral absorber (dashed line) and the ionised components for the 02 spectrum. 
Component 02a is displayed with the 
light line and 02b with the dark line. The most relevant transitions are also labelled. 
Middle panel: \rg\ data together with the best fit. Lower panel: residuals from the best fit. 
} 
\end{figure*}

\begin{figure}
\resizebox{\hsize}{!}{\includegraphics[angle=90]{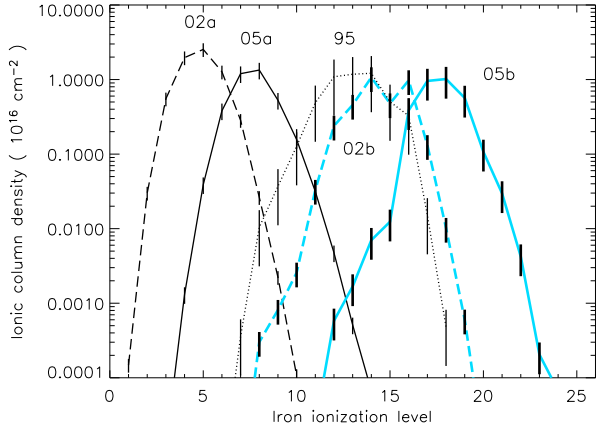}}
\caption{Ionic column density as a function of the ionisation level of iron in the warm absorber components
detected in recent years by \xmm\ (02a, 02b, 05a and 05b) and ASCA (95).}
\label{f:fe}
\end{figure}

\zw\ was observed by \xmm\ for the first time on June 22, 2002 (rev. 0464) for 20\,ks \citep{luigi04}. At the time the source 2--10\,keV flux was about 1.7 times higher than in the present observation and showed remarkable 
spectral complexity at 
low energies. In Fig.~\ref{f:ratio0205} we over-plot the two \epi\ data sets, but with the 2005 spectrum shifted up in order to best match the continuum during 2002. A qualitative inspection of the figure shows that the soft energies
underwent the most dramatic changes between the two epochs (we refer to the 2005 and 2002 data sets as 
05 and 02, respectively).
We reprocessed the 02 \rg\ and \epi\ data using SAS~7.0 and we focused the analysis mostly on the warm absorber. 
The \rg\ data were rebinned to reach a S/N ratio $\sim9$ over most of the band. 
The best fit shows that the underlying continuum changed between 02 and 05 mostly in flux rather than shape,
which is here again described by a simple broken power law (Tab.~\ref{t:rgs_02}). 
The shape of the ionising continuum was again constructed using the unabsorbed \epi\ continuum and the fluxes recorded
in optical/UV by OM (dashed line in Fig.~\ref{f:sed}). In 02 only the OM-UVW2 filter was used.   
Similarly to the 05 data, 
in addition to the neutral gas, we find evidence for two
ionised absorption components. The corresponding 
ionisation parameters are $\log\xi\sim-0.9$ and $\log\xi\sim1.6$. 
We identify these two warm absorber components as 02a and 02b, respectively (Fig.~\ref{f:rgs_tra02}). 
Interestingly, we find evidence that the 02a component outflows with a velocity similar to the 05a component ($v_{\rm
out}\sim1800$\,\kms). The width of the lines is $\sigma=67\pm50$\,\kms.
Going further back into the recent history of the X-ray absorbers of \zw, we reanalysed 
the ASCA data obtained in 1995 \citep{leighly99}. The continuum is once again well fitted by a
broken power law with parameters comparable to the most recent 05 \xmm\ spectra: $\Gamma_{\rm soft}=2.58\pm0.07$,
$\Gamma_{\rm hard}=2.35\pm0.07$ and a break energy at $2.0\pm0.8$\,keV. 
We confirm the presence of a neutral gas phase with column density $N_{\rm H}=4\pm1\times10^{20}$\,cm$^{-2}$,
consistent with the values found with \xmm.\\
For the ASCA data, simultaneous UV observations were unavailable; therefore a true SED could not be constructed. 
The ionising SED was then built using the ASCA continuum plus the same $\alpha_{\rm ox}$
\footnote{
$\alpha_{\rm ox}$ is the spectral index between 2500\,\AA\ and 2\,keV \citep{tan79}.} $=-1.25$ found in 05. We used this SED only to 
obtain a best fit that consistently accounted for the absorption features and then only considered 
the ionic column densities (and not the ionisation parameter) to compare with our \xmm\ results.\\
As the ionisation parameter is dependent on the ionisation balance (i.e. the SED), a good way to compare warm absorbers
produced by different (or incomplete) SEDs, is to evaluate the different elemental ionic column densities. 
In particular, iron is the most
sensitive to changes in the ionisation of the gas \citep[e.g.][]{behar01}. 
In Fig.~\ref{f:fe}, we plot the ionic column densities as a function of
the ionisation stage of iron for each of the X-ray ionisation components detected in this source. 
The column densities are obtained from the XABS best-fit models. 
The relative errors are simply the scaled errors from the total hydrogen column densities
of the XABS ionisation components.
The 02 components are
well separated: 02a peaks at \fev, while 02b peaks at \fexiv-\fexvi. The 05a component has a relatively lower 
total column density (Tab.~\ref{t:rgs_05}) and shares some ions with the 02a components. The 05b component displays the highest ionised iron transitions. The ASCA component
(labelled 95) is more in the high ionisation region.\\       
A possible second component in the ASCA spectrum would be difficult to detect. We verified that 
a higher ionisation component, with a column
density of the order of $10^{21}$\,cm$^{-2}$ would produce mostly lines (see Fig.~\ref{f:rgs_tra05}, \ref{f:rgs_tra02}), which
would be undetectable in the ASCA spectrum. A much less ionised gas peaking, say at \feiii-\fev, would be easily confused
with the neutral absorber. 

\begin{table}
\caption{\label{t:rgs_02}2002 best-fit model parameters for 
RGS and PN. Fluxes are in units of 10$^{-12}$ erg\,s$^{-1}$\,cm$^{-2}$, unabsorbed 
luminosity in
units of 10$^{43}$ erg\,s$^{-1}$. Component 1: broken power law, 2:neutral absorber, 3,4: warm absorbers. 
All emission components are modified by Galactic photoelectric absorption.}
\begin{center}
\begin{tabular}{llll}
\hline
comp. & param. &  R1+R2 & PN\\

1& $\Gamma_1$ 				& 2.8$\pm$0.1 	& 2.71$\pm$0.07\\
 & $\Gamma_2$ 				& $\dots$ 		& 2.32$\pm$0.07\\
 & $E_{\rm break}$\,(keV)			& $\dots$ 		& 1.9$\pm$0.2\\
2 & $N_{\rm H}$ 			&3.0$\pm$1.6		& 4.2$\pm$0.6\\

3 & $N_{\rm H}$ (10$^{20}$ cm$^{-2}$) 	& 24$\pm$5		& 15$\pm$3\\ 
& $\log \xi$ 				& --0.9$\pm$0.2	& --0.8$\pm$0.3\\
& $v_{\rm out}$ (\kms) & --1800$\pm400$ & --1800 fix.\\

4 & $N_{\rm H}$ 			& 13$\pm$4 		& 9$\pm3$\\
& $\log \xi$ 				& 1.6$\pm$0.2 		& 1.3$\pm$0.3\\

\hline
& $\chi^2/\nu$ & 231/211 & 806/788\\

&  F$_{2-10\,{\rm keV}}$&$\dots$ & 7.9$\pm0.26$\\
&  F$_{0.5-2\,{\rm keV}}$ & 8.46$\pm1.34$&  8.1$\pm0.5$  \\
&  L$_{2-10\,{\rm keV}}$ &$\dots$ &7.47$\pm0.25$ \\
&  L$_{0.5-2\,{\rm keV}}$& 1.49$\pm0.22$& 1.26$\pm0.36$ \\
\hline
\end{tabular}
\end{center}
\end{table}

\section{Discussion}

\subsection{The X-ray warm absorber}

\begin{figure}
\resizebox{\hsize}{!}{\includegraphics[angle=90]{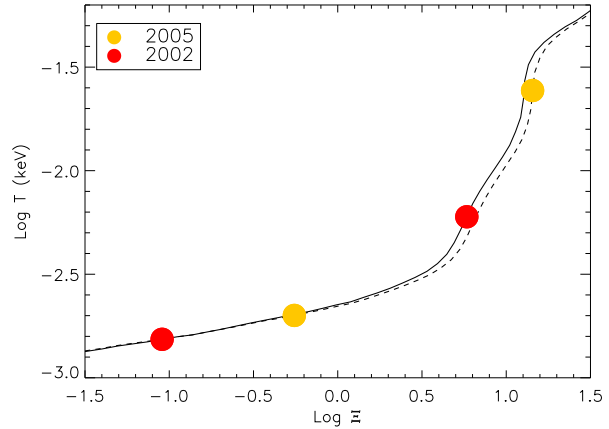}}
\caption{Stability curve for the 2002 (solid line) and 2005 (dashed line) observations. 
The dark circles represent the two ionised
components detected in 2002, while the light circles represents the 2005 components. 
None of the components share the same pressure
ionisation parameter.}
\label{f:xit}
\end{figure}

The 02 and 05 spectra clearly show absorption by ionised gas. 
There were two detected components in 02, with ionisation parameters 
$\log\xi$= --0.9 and 1.6. The 05 spectrum was also well fitted with two 
warm absorber components, showing absorption with higher ionisation ($\log\xi\sim0\ {\rm and}\ 2.6$). 
In Fig.~\ref{f:xit} we plot the gas temperature ($T$) against the ionisation pressure parameter $\Xi$ 
for the 02 and 05 SEDs (drawn with the solid and dashed lines, respectively). The ionisation pressure parameter is defined as
$\Xi\equiv L/4\pi r^2cp$, where $L$ is the ionising luminosity, $r$ the gas distance from the source, and $p$ the gas
pressure. The dark dots mark the position of the 02
warm absorbers on the stability curve relative to the SED at that epoch, while the light dots are 
the components detected in 05. Our curves show a wide interval of $\log\Xi$ in which 
the absorbers exist in a cold regime. This is where the ``cold" components 02a and 05a lie. 
The ascending part ($\log\Xi>0.5$) of the curve does not present the typical turnover S-shape 
\citep[e.g.][]{crenshaw99}. This is because of the relatively steep X-ray SED of \zw\ \citep[][]{krolik81,guilbert}. 
We note that both the two 02 and two 05 components lie far apart from each other 
and are not in pressure equilibrium. This behaviour has been observed in other Seyfert~1 galaxies \citep[e.g.,][]{steen05,elisa06}. This
suggests that conditions other than pressure equilibrium \citep{kk95} should be present to make the two absorbers co-exist, such as magnetic confinement, as already suggested for the broad line region clouds
\citep[e.g.][]{rees87}. The lower ionisation $a$ components show an outflow velocity of roughly 1700-1800\,\kms\ 
in 02 and 05. Therefore there are no strong velocity changes apparent over 
time, although the constraints on velocity changes are limited.

\begin{figure}
\resizebox{\hsize}{!}{\includegraphics[angle=90]{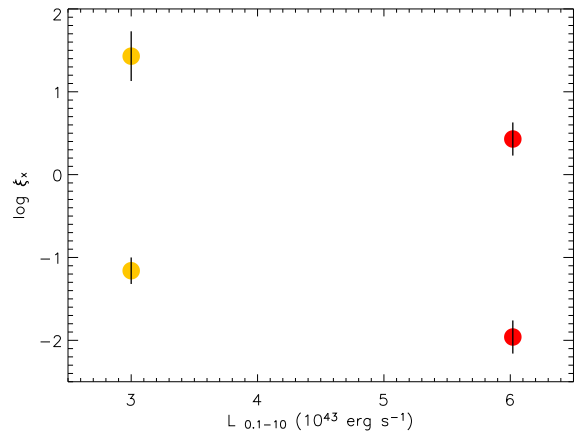}}
\caption{The log of the X-ray ionisation parameter, $\xi_{\rm X}$, against the 0.1--10\,keV intrinsic luminosity for the 02 (dark dots) and
05 (light dots) components.}
\label{f:antic}
\end{figure}

\subsection{Long-term changes in the warm absorbers}

The X-ray ions are 
mostly influenced by the 0.1--10\,keV radiation, rather than by the UV photons \citep[e.g.][]{netzer96}. 
Therefore in the discussion below we shall use the X-ray ionisation parameter $\xi_{\rm X}$, obtained simply by scaling down the ionising luminosity to the 
0.1--10\,keV band \citep[e.g.][]{george98}. 
This calculation can be done only for the \xmm\ data sets, for which the complete ionising SEDs were available. 
The values for $\xi_{\rm X}$ then become $-1.96\pm0.2$ and $0.43\pm0.22$ for the 02a and 02b 
components, respectively. The 05a and 05b values are 
$-1.16\pm0.16$ and $1.43\pm0.3$, respectively. In Fig.~\ref{f:antic}, $\xi_{\rm X}$ of both the 02 and 05 components
are plotted against the 0.1--10\,keV 
luminosity. Note that in general the ionisation parameter value 
is SED dependent. For the 02a and 05a components the corresponding 
stability curves \citep[which are mostly dependent on the SED shape,][]{bottorff} 
are indistinguishable (Fig.~\ref{f:xit}). The stability curves start to differ slightly 
at higher values of $\Xi\propto\xi/T$, therefore we compare the 02b and 05b results with this caveat in mind. 
We note a significant anti-correlation between the 0.1$-$10\,keV luminosity and the X-ray ionisation parameter. 
For a twofold decrease in luminosity between 02 and 05, the low ionisation ($a$ components) $\xi_{\rm X}$ 
{\em increases} by a factor of $\sim6.2$. For the 02b and 05b components, the increase factor is $\sim10$. 
We may consider the $a$ and $b$ absorbers of 02 and 05 to be low- and high-ionisation components. 
If the 
observed ionisation comes from gas in equilibrium, we should invoke either a decrease in density 
(i.e. the gas thickness should increase, as $N_{\rm H}$ did not change dramatically) 
between 02 and 05, or movement of the gas {\it closer} to the central source. 
The latter seems an unlikely possibility as blue-shift of the 
$a$ components is observed in 02 and 05. If the gas consists of discrete outflowing narrow filaments or clouds, 
then we do not expect to see the same structure along our line of sight if the distance $\Delta r$ travelled by this gas is much larger than its characteristic size $d$. 
For a rest-frame time difference of $\sim$1057 days between both observations and the 
measured outflow velocity of 1700~km\,s$^{-1}$, $\Delta r \sim 1.5\times
10^{16}$~cm.\\
For the following calculation we consider the case $\Delta r \gtrsim d$,
i.e. the size of the cloud is comparable to the gas travel path. The
absolute distance of the centre of the cloud $r$ from the source should be much
larger than the sizes considered here, therefore we also assume that the change in
$r$ is not a critical term. We then compute\\

\noindent
\begin{math}
\frac{\xi_{05}}{\xi_{02}}=\frac{L_{05}}{L_{02}}\times\left(\frac{N_{\rm H}}{d}\right)_{02}\ \times\left(\frac{d}{N_{\rm H}}\right)_{05},\\
\end{math}

\noindent
where $nd=N_{\rm H}$. The observed $\xi_{\rm X}$ ratio between 05 and 02 is 7 for the $a$ components. 
The
X-ray luminosity ratio is 0.5. With a simple calculation we obtain $d_{05}/d_{02}=7.6$.
Keeping in mind the assumption $\Delta r> d_{02}\sim 1.5\times 10^{16}$~cm, for the $a$ components we obtain 
$n_{02}<1.6\times10^{5}$\,cm$^{-3}$ and $n_{05}<1.1\times10^{4}$\,cm$^{-3}$. These are very low values 
compared to the estimated limits 
for the warm-absorber densities $2\times10^{6}-10^{9}$\,cm$^{-3}$ \citep{netzer02,kraemer02}, 
therefore it seems unlikely that we are observing the same portion of the gas in both 02 and 05.\\
A scenario that could 
reproduce the observed anti-correlation sees the gas constantly in a non-equilibrium state \citep[e.g.][]{nicastro99} 
because, for example, of a delay in the gas response. 
Such a delay may be due to the gas having a very low density \citep[$10^{7-8}$\,cm$^{-3}$, ][]{nicastro99}. 
Alternatively, the outflowing gas may be ``transient", in the form of clouds or filaments passing through the line of sight. 
A variety of distinct ionisation stages, seemingly unrelated to the ionising flux, could be then 
accounted for. This explanation is strengthened if, contrary to what is observed for \zw,   
components with very
different column densities were observed at different times \citep[e.g.][]{kraemer01,kraemer05,costantini00}.  
The outflow velocity may not change in time, if the gas is also moving 
transversely \citep{crenshaw_review,kraemer05}. 
Finally, the anti-correlation seen in Fig.~\ref{f:antic} may be the result of a shielding of the gas 
which is thus experiencing a filtered ionising luminosity. Qualitatively, the
(self-)shielding of the gas may happen away 
from our line of sight, and is not directly observable in a complex gas environment 
as depicted for example in \citet[][]{proga05,progak04} for the UV and X-ray emitting regions. 

\begin{figure}
\resizebox{\hsize}{!}{\includegraphics[angle=90]{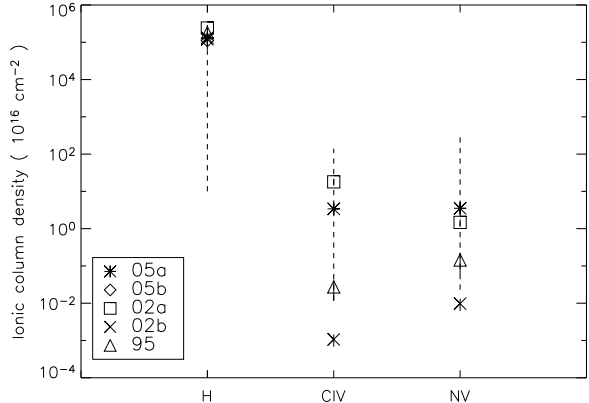}}
\caption{Column density of hydrogen (\hi+\hii), \nv\ and \civ\ predicted by the X-ray absorbers at different epochs. The dashed
vertical lines show the range of $N_{\rm H}$ inferred from the HST measurements taken in 1994 \citep[][]{laor97}. }
\label{f:uv}
\end{figure}

\begin{figure}
\resizebox{\hsize}{!}{\includegraphics[angle=90]{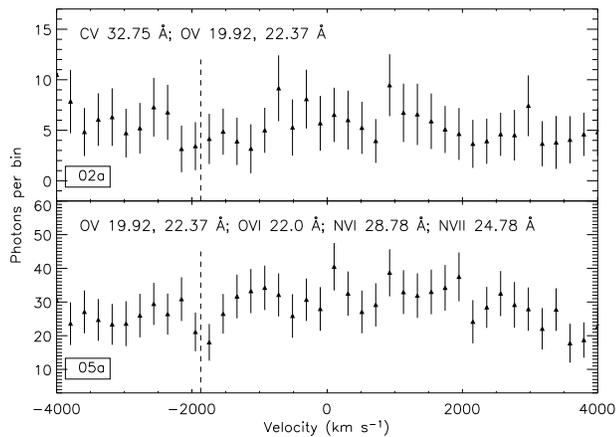}}
\caption{\xmm\ velocity spectra for the 02a (upper panel) and 05a (lower panel) components. The bin size is 200\,\kms. 
The transitions chosen to produce the stack of the (unblended) line profiles for each absorber are also labelled. 
The vertical dashed line marks the blue-shift as measured by HST-FOS. Note that
the redshift used in the HST analysis was 0.0608 which implies a shift in the
outflow velocity of only 90\,\kms, which is negligible compared to the errors
associated to the X-ray measurements.}
\label{f:vel}
\end{figure}

\subsection{The connection with the UV absorber}
Some of the gas components detected in the X-rays might also absorb in the UV band. 
In the HST spectrum of \zw, obtained in
1994 \citep{laor97}, the presence of \nv\ and \civ\ absorption features with an outflow velocity of $-1870$\,\kms, 
clearly highlighted the presence of a warm absorber. Unfortunately the column densities of the absorption troughs 
were poorly determined in the HST spectrum: $\log N_{\rm CIV}=14.07-18.14$ and 
$\log N_{\rm NV}=14.36-18.51$. The derived total hydrogen column density was $\log N_{\rm H}=17-20.77$.
The UV measurements (displayed with the dashed vertical lines) are compared in Fig.~\ref{f:uv} 
with the values obtained in the
X rays at different epochs.
The ionic column densities of \nv\ and \civ, which are visible only in the UV band because of their low ionisation, 
were derived from the best-fit models of the X-ray absorbers. 
The $a$ X-ray components have values of $N_{\rm NV}$ and $N_{\rm CIV}$ which are compatible with the 
UV column densities. Also the ASCA column density measurements are consistent with the UV values, while 
the $b$ components clearly do not produce enough of these low-ionisation ions.
Interestingly, the UV
flux in 1994 was about four times higher than in 02 and 05. This at least suggests
that a low ionisation outflowing component may survive years, despite large changes in the UV flux (and presumably of 
the overall level of the SED). 
The UV total hydrogen column density is slightly lower than for the X-ray
absorbers (Fig.~\ref{f:uv}). 
The outflow velocity measured from the X-ray spectra is consistent with the value found in the HST-FOS spectrum. This is shown in Fig.~\ref{f:vel} where 
the velocity spectra for the 02a and 05a components are compared \citep[see e.g.][for details on the procedure]{kaspi02}. The vertical dashed line marks the blue-shift as measured from the UV \nv\ line \citep[--1870\,\kms;][]{laor97}.  
The outflow velocity of the UV-X-ray gas did not change in time, within uncertainties. 
This suggests that we are observing a gas not dramatically influenced by changes of the central source (e.g. changes in the acceleration of the flow, or strong variability of the source). 
Therefore the observed gas may be in the terminal-velocity portion of the flow, at a considerable 
distance from the central source, for example at a parsec scale, produced either by 
evaporation in a molecular torus \citep{kk01}, or by instabilities of 
the outer rim of the accretion disc, which in principle can extend to large distances.
As a note of caution, 
the comparison between the X-ray and the UV absorbers discussed here is based on 
non-simultaneous data, which may be
misleading \citep{cren03}. In addition,
the UV information given by HST gives only partial information on the SED, which would be important for an
unambiguous comparison between the UV and X-ray bands.

\section{Conclusions}
We have presented the \xmm\ low-energy X-ray data of \zw, collected in 2005. In the absorbed spectrum  
we find evidence of a neutral phase, stable at least
over the last ten years and probably associated with the host galaxy or with the interaction stream to the companion galaxy of
\zw\ (see also Paper~I). In addition, two ionised absorption 
components with similar column densities of $\sim few\times10^{21}$\,cm$^{-2}$ and ionisation parameters
$\log\xi\sim0\
{\rm and}\ 2.5$ are detected. Comparing these results with a previous \xmm\ observation and archival
ASCA and HST data we conclude:
\begin{itemize}
\item The two ionised gas components detected in both 05 and 02 cannot be in pressure equilibrium with each 
other. For the two
components to coexist, a different mechanism, like magnetic confinement and/or gas shielding has to play a role.

\item The X-ray ionisation parameter $\xi_{\rm X}$ is clearly anti-correlated with the X-ray luminosity. 
A viable
explanation for this considers a gas 
organised in filaments crossing the line of sight. However the similarity
of the column densities of the $a$, $b$ components and also of the gas detected in the ASCA spectrum, make this
possibility not as clear as for other cases \citep[e.g. NGC~4151,][]{kraemer05}. Other possibilities, like
a non-equilibrium state of the absorber, would need continuous monitoring to be rigorously tested. Finally, we may think that 
the observed gas might be shielded from the continuum source in the context of a complex and turbulent 
gas environment and experience a filtered ionising flux.

\item The gas producing the X-ray lower ionisation absorbers detected in 02 and 05 can also produce the
low-ionisation species (\nv\ and \civ) seen in the HST spectrum \citep{laor97}, as the column densities 
and outflow velocities are consistent. Interestingly however, the UV absorber was observed when the UV flux was
about 4 times higher than the ones recorded by \xmm-OM. This shows that  
a low ionisation component can survive for at least ten years, despite significant variations in the UV flux. 
The constant outflow velocity of the UV and the X-ray absorbers in time suggests also that we are observing 
the terminal velocity portion of the gas, probably located at a large distance from the source. The caveat is that 
the X-ray-UV comparison is performed with non-simultaneous
observations, which might be misleading.
\end{itemize}

\section{Acknowledgements}
We thank Jelle Kaastra for useful discussion. 
The Space Research Organisation of the Netherlands is supported
financially by NWO, the Netherlands Organisation for Scientific
Research. WNB acknowledges support from NASA LTSA grant NAG5-13035 and NASA grant NNG05GR05G. 
This research has made use of the Tartarus (Version 3.1) database, 
created by Paul O'Neill and Kirpal Nandra at Imperial College London, 
and Jane Turner at NASA/GSFC. Tartarus is supported by funding from PPARC and NASA 
grants NAG5-7385 and NAG5-7067.

 \vfill\eject
\end{document}